\newcommand{\vdate}{August 1996}
\newcommand{\cernnr}{96-219}
\newcommand{\beq}{\begin{equation}}
\newcommand{\eeq}{\end{equation}}
\newcommand{\beqn}{\begin{eqnarray}}
\newcommand{\eeqn}{\end{eqnarray}}

\newcommand{\dgsm}[1]{\noindent{\Large\bf #1}}

\documentstyle[12pt,epsfig]{article}

\newlength{\sectionnumbersize}
\newlength{\sectionsize}
\begin{document}

\begin{titlepage}

\renewcommand{\thefootnote}{\fnsymbol{footnote}}
\setcounter{footnote}{0}

\begin{flushright}
\hfill CERN-TH/\cernnr\\
\hfill hep-th/9608122\\
\end{flushright}

\vspace{0.5cm}

\begin{center}

{\Large\bf
   {Non-Integrability of Two-Dimensional QCD
   }
}

\vspace{1cm}

\vspace{0.5cm}
\vspace{0.5cm}

{\bf Werner Krauth$^1$ {\it and}
Matthias Staudacher$^2$
}

\vspace{2mm}

\vspace{3mm}
{{}$^1 $CNRS-Laboratoire de Physique Statistique de l'ENS\\
24, rue Lhomond\\
F-75231 Paris Cedex 05, France}

\vspace{3mm}
{{}$^2 $CERN, Theory Division\\ CH-1211 Geneva 23, Switzerland}

\vspace{1.0cm}

\footnotetext[0]{Electronic mail addresses:
krauth@physique.ens.fr, matthias@nxth04.cern.ch.}

\begin{abstract}
In this paper we numerically demonstrate that massless
two-dimensional QCD is not integrable. To this aim,
we explicitly solve the 't Hooft integral equation for
bound states by an adaptive spline procedure, and compute
the decay amplitudes. These amplitudes significantly differ
from zero except in all cases in which the decay also
produces a pion.
\end{abstract}

\end{center}

\vfill
\noindent
\begin{minipage}[t]{5cm}
CERN-TH/\cernnr\\
\vdate
\end{minipage}
\vspace{1cm}
\end{titlepage}

\renewcommand{\thefootnote}{\arabic{footnote}}
\setcounter{footnote}{0}

\newpage
\section{Introduction and Conclusions}

Integrability imposes very strong constraints on a quantum field
theory.
The scattering amplitudes become purely elastic and pair production
is suppressed. These features are rather unrealistic from the point
of view of four-dimensional quantum field theories of elementary
particles; and indeed, integrability only exists in two dimensions.
There it is however ubiquitous, and allows in many cases the
determination of the exact S-matrix.

One of the most interesting two-dimensional model quantum field
theories is $QCD_2$. Pure Yang-Mills theory
in two dimensions is essentially trivial due to the absence of
transverse degrees of freedom; however, adding minimally coupled
fermions results in a model which is believed to display some
of the realistic features of the four-dimensional theory.
Indeed, as 't~Hooft \cite{1} first
showed in the leading ${1 \over N}$
approximation ($N$ being the order of the gauge group), the model
contains a tower of mesons built from quark-antiquark pairs, with
nearly linearly rising meson masses. No free quarks are present.
This furnishes
a simple yet explicit example for confinement.

Recently, it was argued in a series of papers that the model
should be integrable in the case of massless bare quarks
\cite{2} for any $N$.
This conclusion was reached from the study of a bosonized
version of the model, in which seemingly
an infinite number of conserved charges existed ({\it cf} \cite{3}
for a review of this and related work).
An ansatz for an exact, factorized S-matrix was proposed in \cite{4}.

The assertions contained in the above work are very strong,
and clearly warrant verification
in a more direct, and controlled fashion. This may be
accomplished by going back to  't~Hooft's large $N$ limit.
Indeed, the large $N$ limit has been a very useful tool to test
the S-matrix of integrable vector-like theories like the
$O(N)$ model or the $U(N)$ Gross-Neveu model \cite{5}.
The crucial idea is to investigate the occurrence of
pair production \cite{6}.

In the following we carefully compute the decay amplitudes in
the chiral 't~Hooft model. This requires a numerical
solution of 't~Hooft's bound state integral equation for which
no general analytic solution
is known to date. We present in this paper an adaptive spline
procedure,
which allows to compute numerically exact solutions (both in the
massless
and the massive case) with negligible effort. The numerical code is
made available to the interested reader; it outperforms the
algorithms which were previously applied to this problem.
The method most frequently used in the past is based on the use
of Chebyshev polynomials and was developed in \cite{7}.

We definitely find non-zero probabilities for pair production to
first
order in $1 \over N$. This result renders inconceivable
the integrability of chiral two-dimensional $QCD$ with
gauge group $SU(N)$.

In contrast,  we establish the complete on-shell decoupling
of the massless $N=\infty$ pion.  While expected on general grounds,
the result displays a mathematically intriguing property
of 't~Hooft's bound state problem. The decoupling is satisfied  to
a very high numerical precision, which illustrates the quality of our
method.

\section{Decay Amplitudes in the 't~Hooft Model}

$QCD_2$ with gauge group $SU(N)$
is described by the Lagrangian
\beq
\label{lag}
{\cal L} = - {N \over 4 g^2}{\rm Tr}~F_{\mu \nu} F^{\mu \nu}
+\bar \psi (i \gamma^\mu D_\mu - m) \psi,
\eeq
where the field strength is
$F_{\mu \nu}=\partial_\mu A_\nu - \partial_\nu A_\mu +
i [A_{\mu},A_{\nu}]$ and the covariant derivative is
$D_\mu=\partial_\mu +i A_\mu$.
The gauge potential $A_\mu$ is given by $N\times N$ hermitian
traceless matrices. In the  light-cone gauge
and the limit $N \rightarrow \infty$, 't~Hooft
has derived the Bethe-Salpeter equation for meson bound states
\cite{1}:
\beq
\label{thooft}
\mu^2~\phi(x)={\gamma -1 \over x(1-x)}~\phi(x) -
\int_{0}^{1} dy {{\cal P} \over (x-y)^2}~\phi(y).
\eeq
Here $\phi(x)$ is a light-cone wave function with
meson mass-square eigenvalue $\mu^2$ and bare quark mass-square
$\gamma$. In the chiral limit one has $\gamma=0$.
${\cal P}$ denotes principal value integration and
all mass-squares are in units of ${g^2 \over \pi}$.

At $N=\infty$ the mesons are trivially stable since decay
processes are suppressed by at least a factor of $1 \over N$.
One therefore needs the  amplitude to order $1 \over N$
for a meson decaying into two less massive mesons.
This was first worked out in \cite{8};
the amplitude ${\cal A}(\omega)$ is
\beq
\label{decay}
{\cal A}(i,f_1,f_2;\omega)=
{1 \over 1- \omega} \int_0^{\omega} dx \phi_i (x)
\phi_{f_1}\big({x \over \omega}\big)
\Phi_{f_2}\big({x - \omega \over 1 - \omega}\big) -
{1 \over \omega} \int_{\omega}^1 dx \phi_i (x)
\Phi_{f_1}\big({x \over \omega}\big)
\phi_{f_2}\big({x - \omega \over 1 - \omega}\big).
\eeq
Here $\phi_i$, $\phi_{f_1}$, $\phi_{f_2}$ are the wave functions
of the initial and first and second final mesons, respectively.
The quark and the antiquark of the initial meson go to the second
and first final meson, respectively.
The quark-antiquark-meson vertex function
$\Phi(x)$ (with $x \notin [0,1]$)
is related to the wave function $\phi(x)$ through
\beq
\label{Phidef}
\Phi(x) = \int_{0}^{1} dy {1 \over (x-y)^2}~\phi(y),
\eeq
and the on-shell values of the kinematic parameter $\omega$ are
\beq
\label{omegdef}
\omega_{\pm}=
{\mu_i^2 + \mu_{f_1}^2 - \mu_{f_2}^2 \mp
\sqrt{(\mu_i^2 + \mu_{f_2}^2 - \mu_{f_1}^2)^2 - 4 \mu_i^2
\mu_{f_2}^2}
\over 2 \mu_i^2},
\eeq
where $\omega_{+}$ and $\omega_{-}$ correspond to a right moving
and left moving final state $f_1$.
Clearly pair production is only kinematically possible if
$\mu_i \geq \mu_{f_1} + \mu_{f_2}$.
In the special case of
identical final states $f_1=f_2=f$,  eq.~(\ref{omegdef}) simplifies
to
\beq
\label{omegid}
\omega_{\pm}=
{1 \over 2} \mp {1 \over 2} \sqrt{1 - 4 {\mu_f^2 \over \mu_i^2}}.
\eeq
Note that these are partial decay amplitudes, which contain all
the dynamical information. To get the full amplitudes entering
the decay probabilities we have to add the partial ones.
Thus we add the amplitudes for the quark of the
initial particle going to the first or second final state.
One then finds for the total on-shell amplitude
${\cal A}=\big( 1 - (-1)^{\sigma_i +
\sigma_{f_1} + \sigma_{f_2}} \big)
\big( {\cal A}(i,f_1,f_2;\omega_{+}) +
{\cal A}(i,f_1,f_2;\omega_{-}) \big)$.
Here $\sigma=+1$ or $\sigma=-1$ for a state with
even or odd parity. Therefore half of the decays
are forbidden due to the parity symmetry. Below we will only
consider allowed decays.

\section{The Numerical Method}

The integral equation eq.~(\ref{thooft}) is too singular to be solved
by
direct discretization. One is practically  forced to introduce a set
of
smooth test functions, for which the
principal value integral in eq.~(\ref{thooft})
can be computed exactly. Besides being smooth,  an ideal test
function
has to be sufficiently  flexible in order not to impose its own
analytical structure onto the solution of the equation. Cubic splines
(which are also much used in computer graphics) provide
an excellent  approximation scheme  for difficult eigenvalue problems
as the present one.  The power of this approach is very often
overlooked.

Our idea is to search the best solution of the 't~Hooft equation
among the  spline functions $\phi(x)$ on the interval
$[0,1]$. The interpolating cubic spline is uniquely defined by its
function values
$\phi_i = \phi(x_i)$ on a set of  node points
$x_i, i=1,N$ with $x_i < x_{i+1}$ (and by the derivatives at $x=0$
and at $x=1$,
{\it cf} \cite{9}).
Standard algorithms \cite{9},\cite{10} allow to compute the function
which,
on the interval $[x_i, x_{i+1}]$, is a third-order polynomial
and which passes smoothly from any interval into the adjacent ones
(intervals are patched together
in such a way that $\phi$ and its first two derivatives are
continuous at
$x_i$).

For a given set of parameters (the values of $\phi_i, i=1,...,N$, the
derivatives $d \phi / d x |_{x=0}$ and $d \phi / d x |_{x=1}$, as
well as
$\mu^2$), the eq.~(\ref{thooft}) will not be
satisfied, but we can choose the parameters such that the discrepancy
between the l.h.s. and the r.h.s. of eq.~(\ref{thooft}) is minimized.
Using a
standard conjugate gradient algorithm for minimization in high
dimensions,
we easily find  different sets of parameters which let this
discrepancy disappear
for all (numerical) intents and purposes. $\mu^2$
is one of the parameters, the minimization thus effectively solves
the
eigenvalue problem.

As presented,  the spline approximation imposes a locally polynomial
structure, which
is inappropriate especially close to $x=0$ and $x=1$, and
particularly so
in the massive case ($\gamma \neq 0$), where the
wave functions are known to behave like
$\phi(x) \sim x^{\beta}$ and $\phi(x) \sim (1-x)^{\beta}$
with $\pi \beta \cot \pi \beta = 1 - \gamma$.
It may seem, therefore,
that very large values of $N$ have to be used in order to get a
satisfactory solution. To avoid this unwanted feature,
we also include  the values of the $x_i$ in our set of parameters,
which the
minimization algorithm strives to optimize. By doing this, we allow
the program to choose large interval sizes where a polynomial
approximation is {\it locally} appropriate, and small sizes
elsewhere.
We usually increase the precision of the approximation by one order
to several
orders of magnitudes at fixed $N$.
$4$ to $5$ significant digits
of the eigenvalues may be obtained with as few as $N=10$ nodes.
\begin{figure}[htbp] \unitlength 1mm
\begin{center}
\begin{picture}(100,100)
\put(-0.3,0){\epsfig{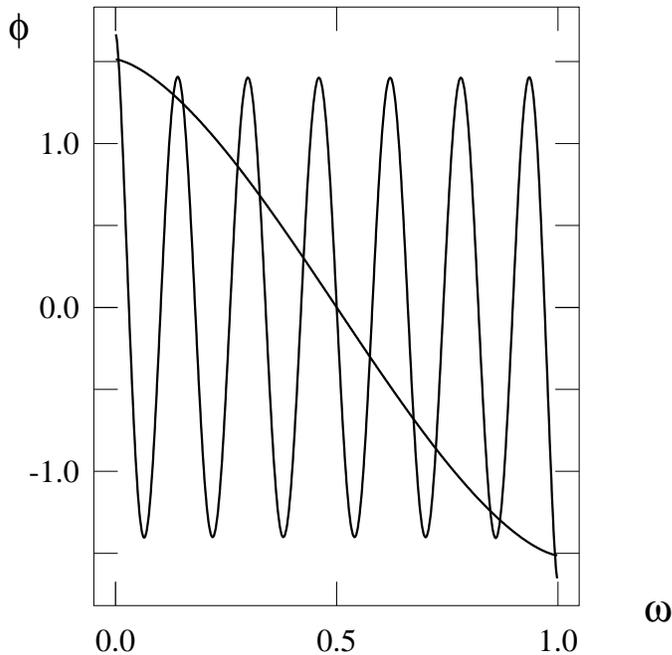}}
\end{picture}
\end{center}
\caption[]
{{\it Superposed first and $13$th excited
state of the massless 't~Hooft equation.}}
\label{figref1}
\end{figure}

In fig.~(1) we give examples of the first and the $13$th excited
state of the massless equation (\ref{thooft}) (all the other states
may be
obtained with the help of the available program). Note that the
ground state
is given by the analytically exact solution
$\phi = 1$. We observe numerically that
the higher excited states states are very well described by cosines,
with
the exception of the endpoints $x=0$ and $x=1$
and a slight $x$-dependent phase-shift when moving away from the
midpoint
$x={1 \over 2}$.

\section{Non-Integrability}

We have computed the  amplitudes for all decays involving states up
to
the 14th excited state and repeated this procedure
for the massive case. The on-shell amplitudes,
while clearly non-zero in general,
display a number of
features that would be interesting to explain. Some of
these features differ significantly between the massless
and the massive case. For example, the amplitude for the decay
$(i,f_1,f_2)$
involving a highly excited state $i >>1$ going to two low-lying
states
$f_1$ and $f_2$ goes -- for increasing $i$  --
rapidly to zero in the massless case while it reaches a maximum
in the massive case.

Examples of our calculations can be found in
fig.~(2) where we display the decay amplitude
${\cal A}(i,f_1,f_2;\omega) $ {\it vs} $\omega$, for different
initial
and final states. In figs.~(2a) and (2b)  we consider symmetric
decays
$(i,f,f)$
in which the on-shell parameters $\omega_{\pm}$ satisfy $\omega_{+} +
\omega_{-} =1$, and where the two decay amplitudes are trivially
identical.

\begin{figure}[htbp]
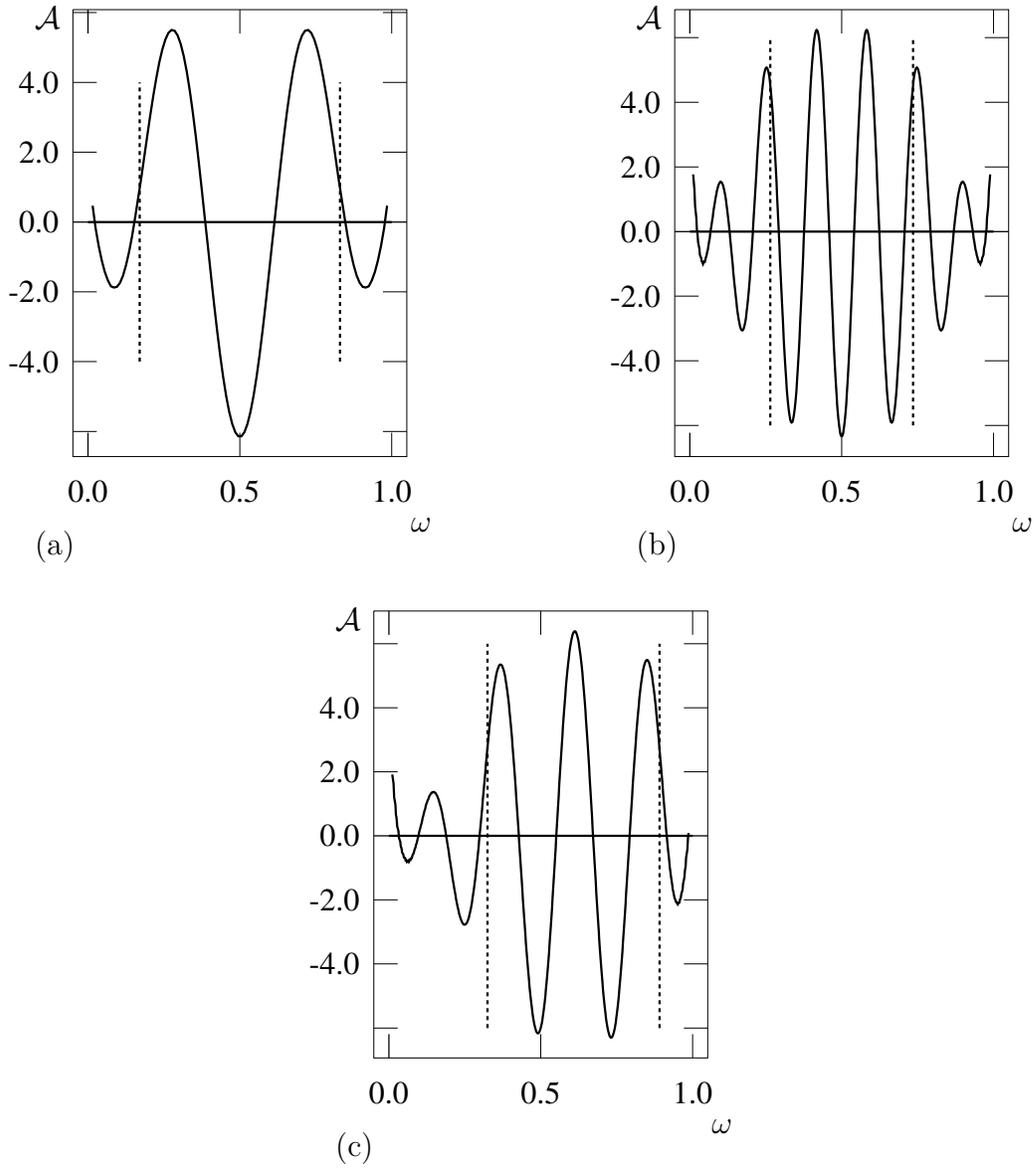
 \unitlength 1mm
\begin{center}
\begin{picture}(160,160)
\put(0,85){\epsfig{file=fig2a.eps,width=8cm}}
\put(80,85){\epsfig{file=fig2b.eps,width=8cm}}
\put(40,5){\epsfig{file=fig2c.eps,width=8cm}}
\put(15,80){\rm (a)}
\put(95,80){\rm (b)}
\put(55,0){\rm (c)}
\put(15,150){${\cal A}$}
\put(65,83){$\omega$}
\put(95,150){${\cal A}$}
\put(145,83){$\omega$}
\put(55,70){${\cal A}$}
\put(105,3){$\omega$}
\end{picture}
\end{center}
\caption[]
{{\it
Amplitudes for three sample decay processes:
(a) (5-1-1), (b) (13-3-3) and (c) (9-3-1).
The vertical bars mark the on-shell values
$\omega_{+}$ (right decays) and $\omega_{-}$
(left decays).
}}
\end{figure}

Note that all these decays are parity-allowed and that we obtain
${\cal A}(\omega_{+}) = {\cal A}(\omega_{-})$. We have carefully
checked the numerics, and
clearly obtain non-zero decay amplitudes proving our
claim: $QCD_2$ is {\it not} integrable.

It should also be mentioned that the first few possible decay
amplitudes are rather small (see e.g.~(5-1-1)).
Fig.~(2b)~(for the decay (13-3-3)) shows
that some decays nevertheless have big amplitudes. This seems to
be the case whenever we are close to threshold (i.e.~when
the momentum of the decay products is small in the rest frame
of the decaying particle.
Finally, we show in fig.~(2c) a case of an asymmetric decay (9-3-1),
in which eq.~(\ref{omegdef}) gives two non-trivially related
numbers for the on-shell
value of $\omega$. Nevertheless, we find that the decay probabilities
are equal to very high precision,
illustrating again the quality of the
numerical solution. Let us mention that the same features are found
in the massive case, which is not further discussed here.
For the parity-forbidden decays
we find non-zero  probabilities of equal magnitude,
but opposite sign (for any value of the mass parameter $\gamma$):
\beq
\label{convol}
{\cal A}(i,f_1,f_2;\omega_{+}) =
(-1)^{1 + \sigma_i + \sigma_{f_1} + \sigma_{f_2}}
{\cal A}(i,f_1,f_2;\omega_{-}),
\eeq
which is a convolution identity that is non-trivial
except for identical final states $f_1=f_2$ (where
$\omega_{+} =1 - \omega_{-}$).
It is expected on physical grounds: The probability
for $f_1$ going to the right or the left has to be the same.

\section{Pion Decoupling}

An intriguing by-product of the present work is the
``experimental'' fact that the pions cannot be produced
through the decay of massive states: ${\cal A} = 0$ as soon
as $f_1$ or $f_2$ are pions.
The existence of massless pions in the 't Hooft model is a
consequence
of chiral $U(1)$ symmetry as shown in \cite{11}.
Although in (\ref{lag}) both $SU(N)$ and $U(1)$ vector symmetry are
gauged,
the large $N$ limit takes the $U(1)$ gauge coupling to zero in a way
that suppresses the chiral $U(1)$ anomaly.
The simplest argument for this is that the anomaly receives
contributions
from Feynman diagrams with one fermion loop, which by standard large
$N$
counting is $O(1/N)$.
Conservation of the vector $U(1)$ current in two dimensions allows
writing
$J^\mu = \epsilon^{\mu\nu}\partial_\nu\phi$, and
conservation of the chiral $U(1)$ current
$J_\mu^5=\epsilon_{\mu\nu}J^\nu$
then implies the existence of a free massless boson \cite{12}.
Thus the pion can be considered an $N=\infty$ artifact, and the above
argument gives an indirect explanation for the observed on-shell
decoupling.

However, it is not at all clear in which way the decoupling
manifests itself directly in the mathematical description of the
't~Hooft model. The pion wave function $\phi_{\pi}$
is the only known exact solution
of the chiral ($\gamma =0$) bound state equation (\ref{thooft}):
\beq
\label{pion}
\phi_{\pi}(x)=1 \;\;\;\; {\rm with} \;\;\;\; \mu_{\pi}^2=0
\;\;\;\; {\rm and} \;\;\;\; \Phi_{\pi}(x)={-1 \over x(1-x)}.
\eeq
If, say, the second final state is a pion ($f_2=\pi$) we
thus expect the amplitude (\ref{decay}) to be zero for any initial
state
$i$ and first final state $f_1$ with $\mu_i > \mu_{f_1}$:
\beq
\label{pidecay}
{\cal A}(\omega_{+})= \int_0^{\omega_{+}} dx~\phi_i (x)
\phi_{f_1}\big({x \over \omega_{+}}\big)
\big[{1 \over \omega_{+} - x} + {1 \over x -1} \big] -
{1 \over \omega_{+}} \int_{\omega_{+}}^1 dx~\phi_i (x)
\Phi_{f_1}\big({x \over \omega_{+}}\big)=0,
\eeq
if we choose the on-shell value (\ref{omegdef}):
\beq
\label{ompion}
\omega_{+} = {\mu_{f_1}^2 \over \mu_i^2}.
\eeq
We indeed verified the validity of eq.~(\ref{pidecay}) for all
possible decays
of the first $14$ excited states, using our numerical method.
Examples are shown in fig.~(3a) (for the decay (5-4-0)) and fig.~(3b)
(for the decay (13-3-0)). In the last example, the  numerically
computed
on-shell value $\omega$ differs from the zero of the function
${\cal A}(\omega)$ by about $10^{-4}$, which is a very good precision
considering the order of the initial excited state, and the small
number of parameters in our spline approximation ($N = 60$ was used).

\begin{figure}[htbp]
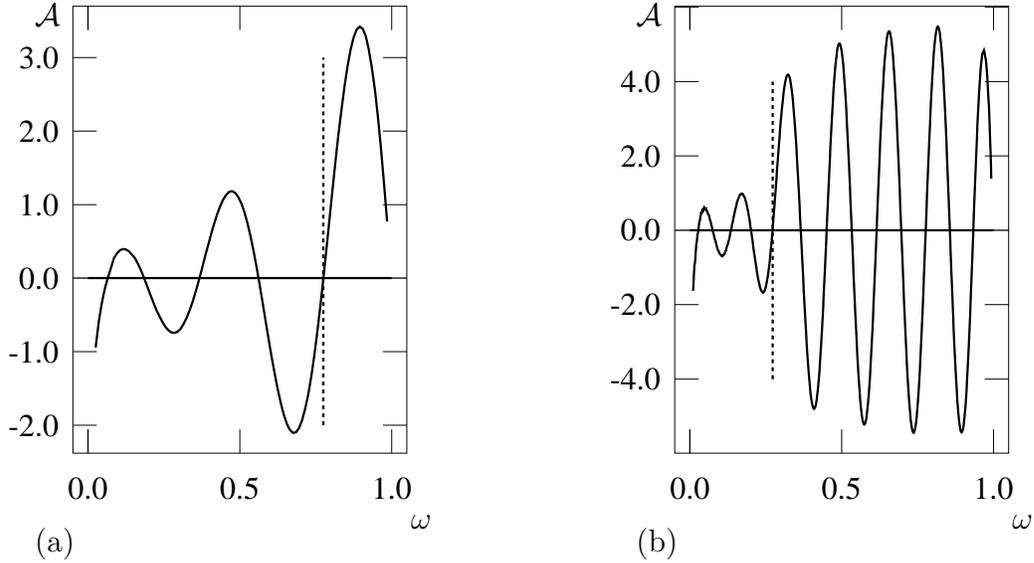
 \unitlength 1mm
\begin{center}
\begin{picture}(160,79)
\put(0,5){\epsfig{file=fig3a.eps,width=8cm}}
\put(80,5){\epsfig{file=fig3b.eps,width=8cm}}
\put(15,0){\rm (a)}
\put(95,0){\rm (b)}
\put(15,70){${\cal A}$}
\put(65,3){$\omega$}
\put(95,70){${\cal A}$}
\put(145,3){$\omega$}
\end{picture}
\end{center}
\caption[]
{{\it
Amplitudes for two sample decay processes, involving
pions: (a) (5-4-0) and (b) (13-4-0). The vertical bar marks
the on-shell value $\omega_{+}$. It is seen
that the pions decouple on-shell.
}}
\end{figure}

We checked that the amplitudes become non-zero as soon
as the quarks acquire a mass.
It is a challenging problem to give a mathematical proof of
the convolution formula (\ref{pidecay}),
starting from the bound state equation (\ref{thooft}).

\vspace{5mm}
\dgsm{Acknowledgements}

\noindent
We would like to thank M. C. Batoni-Abdalla and E. Abdalla
for collaboration in the initial stage of this project,
as well as for several constructive discussions.
W. K. thanks the CERN theory division for hospitality. FORTRAN
programs
can be obtained by e-mail (from any of the authors).


\newcommand{\scs}{\rm}
\newcommand{\bibitema}[1]{\bibitem[#1]{#1}}
\newcommand{\bibbeginlong}{\begin{thebibliography}{XXX\,1988\,\,}}
\newcommand{\bibbeginshort}{\begin{thebibliography}{XX}}

\bibbeginshort 

\addcontentsline{toc}{section}{References}

\bibitema{1}
      G. 't~Hooft, Nucl. Phys. B75 (1974) 461.

\bibitema{2}
      E. Abdalla and M. C. B. Abdalla,
      Int. J. Mod. Phys. A10 (1995) 1611;
      E. Abdalla and M. C. B. Abdalla,
      Phys. Lett. B337 (1994) 347;
      E. Abdalla,
      {\it Towards a solution of two-dimensional QCD},
      hepth-9510139, to be published
      in the proceedings of ICTP Workshop on Strings,
      Gravity and Related Topics, Trieste, Italy, 29-30 June 1995.

\bibitema{3}
      E. Abdalla and M. C. B. Abdalla,
      {\it Updating $QCD_2$},
      Phys. Rept. 265 (1996) 253.

\bibitema{4}
      E. Abdalla and M. C. B. Abdalla,
      Phys. Rev. D52 (1995) 6660.

\bibitema{5}
      A. B. Zamolodchikov and Al. B. Zamolodchikov,
      Ann. Phys. 120 (1979) 253.

\bibitema{6}
      E. Abdalla (unpublished) pointed out to us that
      the integrability of $QCD_2$ could be tested
      by computing pair-production probabilities
      in the 't Hooft model.

\bibitema{7}
      J. Hanson, R. D. Peccei and M. K. Prasad, Nucl.~Phys.~B121
      (1977) 477;
      R. L. Jaffe and P. F. Mende, Nucl.~Phys.~B369 (1992) 189.

\bibitema{8}
      C. G. Callan Jr., N.~Coote, and D.~J.~Gross,
      Phys. Rev. D13 (1976) 1649.

\bibitema{9}
      J. Stoer and R. Bulirsch,
      {\it Introduction to Numerical Analysis},
      Springer Verlag, New York (1980).

\bibitema{10}
      W. H. Press, S. A. Teukolsky, W. T. Vetterling, B. P.
      Flannery, {\sl Numerical Recipes}, 2nd edition,
      Cambridge University Press (1992).

\bibitema{11}
      I. Affleck, Nucl. Phys. B265 [FS15] (1986) 448.

\bibitema{12}
      E. Witten, Nucl. Phys. B145 (1978) 110.

\end{thebibliography}

\end{document}